# Pixelated high-Q metasurfaces for in-situ biospectroscopy and AI-enabled classification of lipid membrane photoswitching dynamics

Martin Barkey[1,§], Rebecca Büchner[1,2,§], Alwin Wester[1], Stefanie D. Pritzl[3,4], Maksim Makarenko[5], Qizhou Wang[5], Thomas Weber[1], Dirk Trauner[6], Stefan A. Maier[1,7,8], Andrea Fratalocchi[5], Theobald Lohmüller[3], and Andreas Tittl[1,*]

[§]equal contribution, *email: andreas.tittl@physik.uni-muenchen.de

[1]Chair in Hybrid Nanosystems, Nano-Institute Munich, Faculty of Physics, Ludwig-Maximilians-Universtität München, Königinstraße 10, 80539 München, Germany

[2]Nanophotonic Systems Laboratory, ETH Zürich, 8092 Zürich, Switzerland

[3]Chair for Photonics and Optoelectronics, Nano-Institute Munich, Faculty of Physics, Ludwig-Maximilians-Universtität München, Königinstraße 10, 80539 München, Germany

[4]Department of Physics and Debye Institute for Nanomaterials Science, Utrecht University, Princetonplein 1, 3584 CC Utrecht, The Netherlands

[5]PRIMALIGHT, Faculty of Electrical Engineering, King Abdullah University of Science and Technology (KAUST), Thuwal 23955-6900, Saudi Arabia

[6]Department of Chemistry, University of Pennsylvania, Philadelphia, PA 19104-6323, United States

[7]School of Physics and Astronomy, Monash University, Wellington Rd, Clayton VIC 3800, Australia

[8]The Blackett Laboratory, Department of Physics, Imperial College London, London, SW7 2AZ, United Kingdom

**Abstract**

Nanophotonic devices excel at confining light into intense hot spots of the electromagnetic near fields, creating unprecedented opportunities for light-matter coupling and surface-enhanced sensing. Recently, all-dielectric metasurfaces with ultrasharp resonances enabled by photonic bound states in the continuum have unlocked new functionalities for surface-enhanced biospectroscopy by precisely targeting and reading out molecular absorption signatures of diverse molecular systems. However, BIC-driven molecular spectroscopy has so far focused on endpoint measurements in dry conditions, neglecting the crucial interaction dynamics of biological systems. Here, we combine the advantages of pixelated all-dielectric metasurfaces with deep learning-enabled feature extraction and prediction to realize an integrated optofluidic platform for time-resolved in-situ biospectroscopy. Our approach harnesses high-Q metasurfaces specifically designed for operation in a lossy aqueous environment together with advanced spectral sampling techniques to temporally resolve the dynamic behavior of photoswitchable lipid membranes. Enabled by a software convolutional neural network, we further demonstrate the real-time classification of the characteristic *cis* and *trans* membrane conformations with 98% accuracy. Our synergistic sensing platform incorporating metasurfaces, optofluidics, and deep learning opens exciting possibilities for studying multi-molecular biological systems, ranging from the behavior of transmembrane proteins to the dynamic processes associated with cellular communication.

**Introduction**

Molecular spectroscopy in the mid-infrared (mid-IR) is an essential tool for studying the structure of complex molecules[1–3]. It probes the characteristic vibrational absorption bands of molecules in this spectral range – known as the 'molecular fingerprint' – and provides unique information about their constituent chemical bonds. However, due to the size difference between micrometer-scale mid-IR wavelengths and nanometer-scale biomolecules, the detection of small quantities of analytes in mid-IR spectroscopy remains challenging. Nanophotonics can bridge this gap in length scales by employing resonant nanostructures, which provide strong and spatially localized near-field enhancements to boost light-matter interactions for increased sensitivity[4,5]. This approach is known as surface-enhanced infrared absorption spectroscopy (SEIRA) and has enabled a variety of sensing applications in fields ranging from biomedicine and pharmacy to food and materials sciences[6–9]. Traditionally,

SEIRA has been realized by using plasmonic resonators. However, due to intrinsic damping caused by Ohmic losses, plasmonic systems are fundamentally limited to comparatively broad resonances with low quality (Q) factors (defined as resonance position divided by line width). The performance of SEIRA approaches can be significantly improved by moving to new materials, such as dielectrics which have low optical losses and high refractive indices[10–13].

All-dielectric metasurfaces supporting bound states in the continuum (BICs) with ultra-sharp resonances have gained broad attention for tailored light-matter coupling applications[14–16]. Conceptually, BIC realizations span the gamut from localized supercavity modes in individual resonators, such as dielectric disks, to extended symmetric-protected modes in metasurfaces[17–19]. Symmetry-protected BIC-driven metasurfaces in particular enable precise control over resonance position, line width, and magnitude of near-field enhancement via the scaling and asymmetry of their unit cells, which makes them ideally suited for the molecular spectroscopy of analytes ranging from biological materials to environmental pollutants[20–22]. BIC-driven metasurface concepts can provide powerful new functionalities when implemented in a pixelated arrangement, where multiple high-Q metapixels are arranged in a two-dimensional array with linearly varying resonance frequency[23], enabling unambiguous mapping between spectral information (i.e., the metapixel resonance wavelength) and spatial information (i.e., the location of the metapixel within the array).

Coating such a metasurface array with an analyte leads to a strong modulation of the resonances of individual metapixels correlated to the absorption bands of the target molecules, allowing the imaging-based readout of biomolecular fingerprints. This molecular barcoding approach has been demonstrated for the detection of absorption signatures associated with simple molecules such as proteins or polymers[24], but has so far not been applied to complex and dynamic molecular biosystems. Additionally, BIC-driven all-dielectric metasurfaces have mostly focused on measurements in dry conditions. These aspects significantly limit the practical applicability of such methods, especially in biology, where molecular dynamics and interactions are ideally studied in their natural, usually aqueous, environment and analytes are often part of a larger molecular background matrix resulting in complicated spectroscopic data.

A notable example of dynamic and complex biological system is the cell membrane, where a multitude of functional biomolecules are embedded in a fluid bilayer membrane comprised of amphipathic lipid molecules. Lipid membranes were traditionally considered to assume the primary role as a functional barrier, owing to their selective permeability for ions or large molecules. In recent years, however, striking evidence emerged that cellular functions and metabolic action are also linked to lipid composition and lipid-protein interactions. Yet, many details on how transient changes of membrane properties or compositions influence cellular functions are still poorly understood.

A platform technology capable of recapitulating the dynamic properties of cell membranes are supported lipid bilayers (SLBs)[26]. Lipid molecules can be assembled on solid supports in such a way that they form a continuous bilayer while maintaining a high degree of lateral mobility. An SLB can be further labelled with proteins or other biomolecules, which are then able to reorganize naturally in this fluid matrix[27]. Controlling lateral fluidity in SLBs typically involves adjustments of the lipid composition[28] or of experimental parameters, which is often not physiological and non-reversible. In this regard, synthetic photoswitchable phospholipids, or photolipids, have emerged as a research tool to reversibly alter and control a variety of SLB properties[29], such as fluidity and thickness[30], lipid order and domain formation[30–35], as well as protein molecular dynamics[36] and photoactivation of mechanosensitive channels[37] by photo-isomerization. Recent studies have further shown the potential of photolipids to trigger the release of molecular cargo from liposomes[32] and lipid nanoparticles[38] and to control protein secretion in living cells by means of light[39]. These examples emphasize the wider applicability of photolipids as molecular nanoagents to emulate membrane function. However, harnessing their full potential to control cellular processes and membrane proteins requires a detailed and quantitative understanding of the photoisomerization dynamics and their interactions with other membrane components in a bilayer setting.

Vibrational spectroscopy is ideally suited to study and characterize the structural and conformational properties of photolipid assemblies without the requirement of an additional spectroscopic label[40]. For example, vibrational sum-frequency generation spectroscopy has been used to gain insights in the molecular ordering of azobenzene-based photolipids assembled in a monolayer on water[41]. Recently, FTIR spectroscopy of photolipid bilayer nanodiscs has likewise been utilized to investigate photolipid isomerization[37]. These examples demonstrate the great potential of IR spectroscopy for membrane studies but also expose the experimental constraints that one faces with most established methods: Far-field approaches do not provide a sufficient resolution to address local membrane heterogeneity. Near-field methods, such as IR scattering scanning near-field optical microscopy[42], enable localized studies but fall short of addressing long-range dynamic bilayer properties. Experimental methods to study photoisomerization dynamics of extended photolipid SLBs label free and *in situ* are thus highly demanded.

But even with the right sensor technology, the sheer volume of data generated by spectroscopic techniques presents a formidable challenge in its interpretation and conversion into actionable insights[43]. Artificial intelligence (AI), with its capacity to handle and analyze large datasets, offers an opportunity to significantly advance biospectroscopy. Notably, AI's ability to identify patterns and relationships in complex datasets is especially valuable in hyperspectral imaging, a technique providing detailed chemical and physical information by capturing and analyzing light across a wide wavelength range[44,45].

Combining hyperspectral imaging, metasurfaces, and AI can yield unprecedented levels of biological insights, potentially catalyzing the development of advanced diagnostics and treatments. Inspired by the idea of Explainable AI (XAI)[46,47], which utilizes automated attribution algorithm to reveal the importance on the input features, we introduce an integrated platform for ultrasensitive in-situ biospectroscopy, which combines an all-dielectric pixelated metasurface with AI feature selection for molecular discrimination and leverage it for resolving the intricate switching dynamics of photoswitchable lipid membranes.

Our specifically engineered BIC-driven metasurfaces overcome longstanding limitations of established IR spectroscopy approaches related to low surface sensitivity and the detrimental effects of background water absorption, enabling the study of complex photoswitching processes involving different membrane conformational states in an aqueous environment. Specifically, we investigate SLBs of azobenzene containing phosphatidylcholine lipids (AzoPC), that can be switched between their *trans* and *cis* form using UV and blue light, respectively[31].

To resolve the minute spectral absorption variations between AzoPC isomers within a lipid bilayer, we employ advanced sampling techniques such as pixel doubling and randomization of pixel order. This improves detection efficiency and remove bias from spatially varying imaging performance or sample distribution across the metasurface. Probing the dynamics of the system through continuous time-resolved measurements, we demonstrate the detection of both the membrane formation as well as the membrane photoswitching with high sensitivity.

In such dynamic microfluidic biospectroscopy measurements, small vibrational signals of interest are often obscured by external influences such as variations in illumination, drifts related to evaporation effects through the microfluidics, or thermal variations. To address these issues and allow for automatic and fast classification of the membrane states, we further implement a feature extraction process guided by explainable AI techniques to reduce the parameters for a convolutional neural network (CNN). It is able to effectively identify patterns and relationships within the spectroscopic data, allowing for more accurate and efficient analysis of the underlying processes. In addition, our one-dimensional CNN architecture enables the use of smaller, more efficient filters compared to 2D CNNs that are broadly applied to computer vision applications, which can improve the speed and performance of the network[48].

Significantly, we combine our machine learning model with cutting-edge explainability methods to gain further insights into the internal workings of our model and understand its decision-making process. Such insights enable the identification and correction of errors in the model and provide important design guidelines for the underlying metasurfaces by quantifying the impact of individual pixels on the classification accuracy. Our AI-enabled and metasurface-driven optofluidic platform can be straightforwardly extended to other molecular detection tasks by adapting the metasurface chips based on the CNN model results, paving the way for automatic and robust molecular classification in multi-analyte systems related to environmental monitoring, photo- and electrocatalysis, or medical diagnostics.

**Results**

The surface-enhanced vibrational spectroscopy element of our platform consists of an all-dielectric pixelated metasurface chip optimized for the mid-IR spectral range and for operation in an aqueous environment, incorporated in a polydimethylsiloxane (PDMS) microfluidic cell (Fig. 1a). The integrated fluidic concept allows for the injection of AzoPC vesicles, which subsequently form a supported lipid bilayer on the metasurface chip through vesicle fusion (Fig. 1b). AzoPC undergoes reversible *trans*-*cis* photoisomerization by illumination at wavelengths of 365 nm (labeled "UV", initiating the *trans* to *cis* transition) and 465 nm (labeled "VIS", initiating the *cis* to *trans* transition), respectively[33] (Fig. 1c).

The changes between the two photoswitching states are directly linked to their molecular absorption spectra as conceptually shown in Fig. 1d. To resolve the minute absorption changes associated with the state change, a pixelated metasurface was implemented with resonances covering the spectral range from 1400 $cm^{-1}$ to 1800 $cm^{-1}$. When coupling to the AzoPC molecular vibrations, the metapixel resonances are attenuated, allowing the retrieval of the absorption signatures from the envelope of the reflectance spectra (Fig. 1d, bottom). Precise differentiation and classification of the closely related *cis* and *trans* states were then achieved using processing of the time-resolved spectroscopic data sets through a CNN-based deep learning model (Fig. 1e).

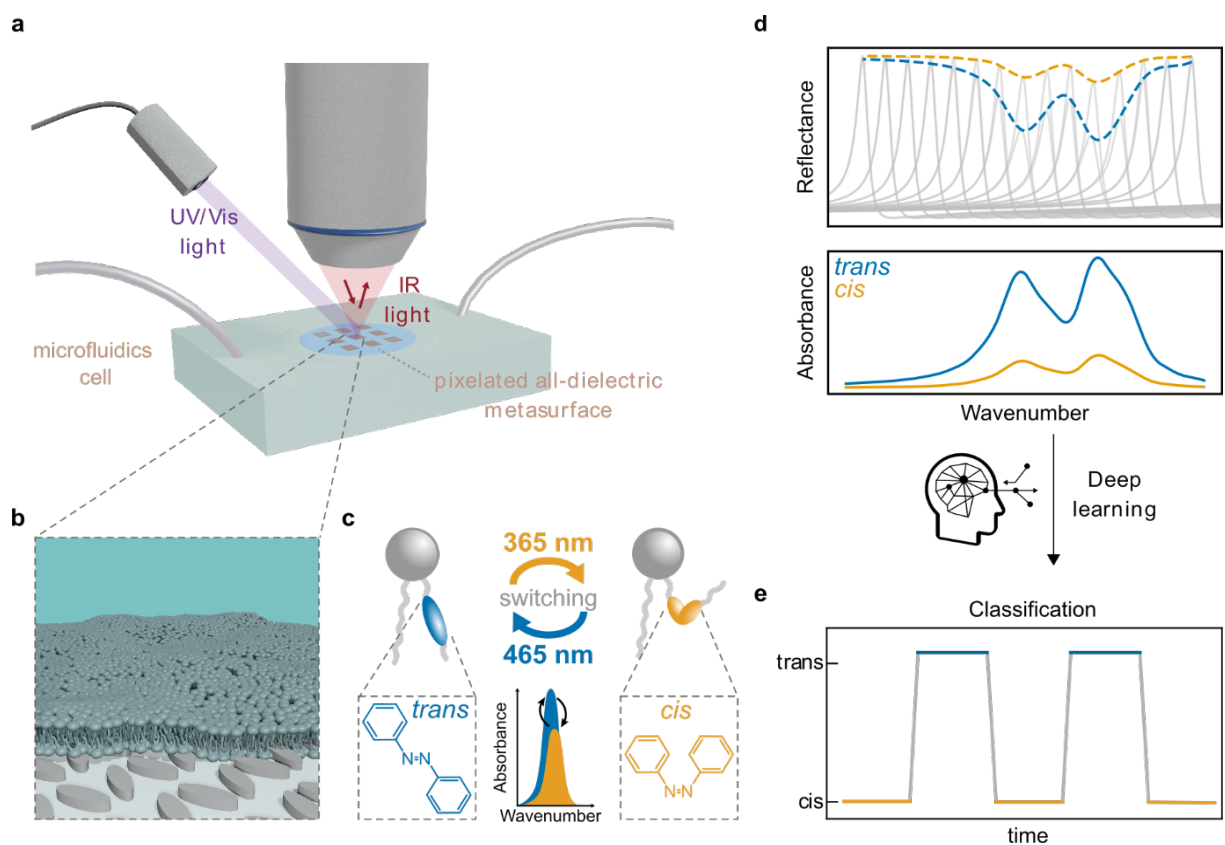

**Figure 1. Metasurface-enabled biospectroscopy aided by AI.** (**a**) Sketch of the metasurface-driven optofluidic biospectroscopy system. A substrate with a pixelated metasurface is integrated into a microfluidics chip, interrogated in reflection with an IR objective, and illuminated with UV/VIS light for photoswitching. (**b**) Sketch of a lipid bilayer on the metasurface highlighting the conformal coating. (**c**) Sketch of the photoinduced change of the lipid azobenzene group in their tails between the *cis* and *trans* conformation upon exposure to UV or visible light. (**d**) Sketch of reflectance spectra of a pixelated metasurface coated with a *cis* and *trans* lipid membrane (bottom) and retrieved absorbance spectra of the lipid membranes (top). (**e**) Classification of the state of the membrane obtained with the deep learning model.

Implementation of the metasurface-based sensor platform started with the numerical design of a 7 by 7 pattern of metapixels with linearly varying resonance positions in the target spectral range (Figure 2a). In the design process, special consideration must be given to the refractive index and absorptive properties of the surrounding medium, $D_2O$, in order to precisely target the resonance position within the desired range of 1400 cm$^{-1}$ to 1800 cm$^{-1}$, where the absorption bands of interest for the AzoPC lipids are located. The unit cell design was optimized for the best trade-off between Q-factor and resonance amplitude in this aqueous environment. Each individual metapixel consists of a periodic array of amorphous silicon (a-Si) ellipse pairs on a CaF$_2$ substrate (Figure 2e,f), where lateral scaling by a factor $S$ is utilized to tailor the resonance position, and the tilting angle $\theta$ determines the asymmetry of the structure and consequently the Q factor[15]. Based on numerical simulations, and ellipse pair design with a thickness of $t = 700\ nm$ and a tilting angle of $\theta = 20°$ was chosen, optimizing the trade-off between Q-factor and near-field enhancement[24].

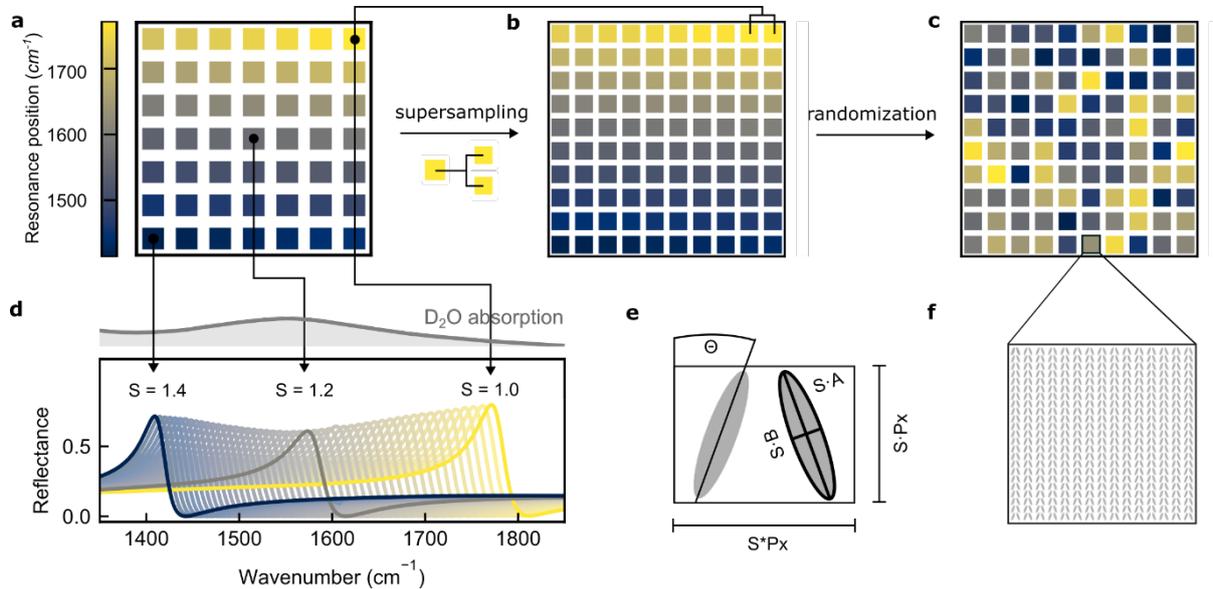

**Figure 2. Metasurface concept and numerical design.** (**a**) Target pixelated barcode with 49 pixels. (**b**) Each Pixel from target barcode is doubled. (**c**) All 98 pixel are ordered randomly revealing the physical metasurface barcode. (**d**) Simulated reflectance spectra of all pixels from target barcode and D2O absorption used in simulations (top inset). (**e**) Sketch of an individual unit cell consisting of a tilted a-Si ellipse pair. (**f**) Sketch of the array of unit cells in one pixel.

Excited by normally incident linearly polarized light, these optimized metapixel designs provide sharp, spectrally clean, and geometrically tunable BIC-driven resonances as shown in Figure 2d. Note that the observed attenuation of the spectra is due to the intrinsic absorption of D$_2$O, the medium used in our subsequent microfluidic measurements. To increase the signal-to-noise-ratio of the chip and to improve its robustness towards spatial variations in the membrane sample, each metapixel was duplicated in a process called pixel doubling to yield a total of 98 pixels (Fig. 2b), which were then arranged randomly throughout a 10 by 10 pixel pattern to avoid any systematic bias induced by the linear resonance scaling (Fig. 2c).Metasurfaces incorporating the randomized array of a-Si metapixels were fabricated on IR transparent $CaF_2$ substrates via a multistep process involving electron beam lithography (EBL) and directional reactive ion etching (RIE) (for details see Methods), with a total structured area on the order of 2 $mm^2$ for each 10 by 10 metapixel pattern (Fig. 3a). Scanning electron microscopy (SEM) images confirm the accurate reproduction of the periodic array structure (Fig. 3b) as well as the target geometrical dimensions of the individual unit cells (Fig. 3c) and dimensions of our tilted ellipses unit-cell geometry (Fig. 3c).

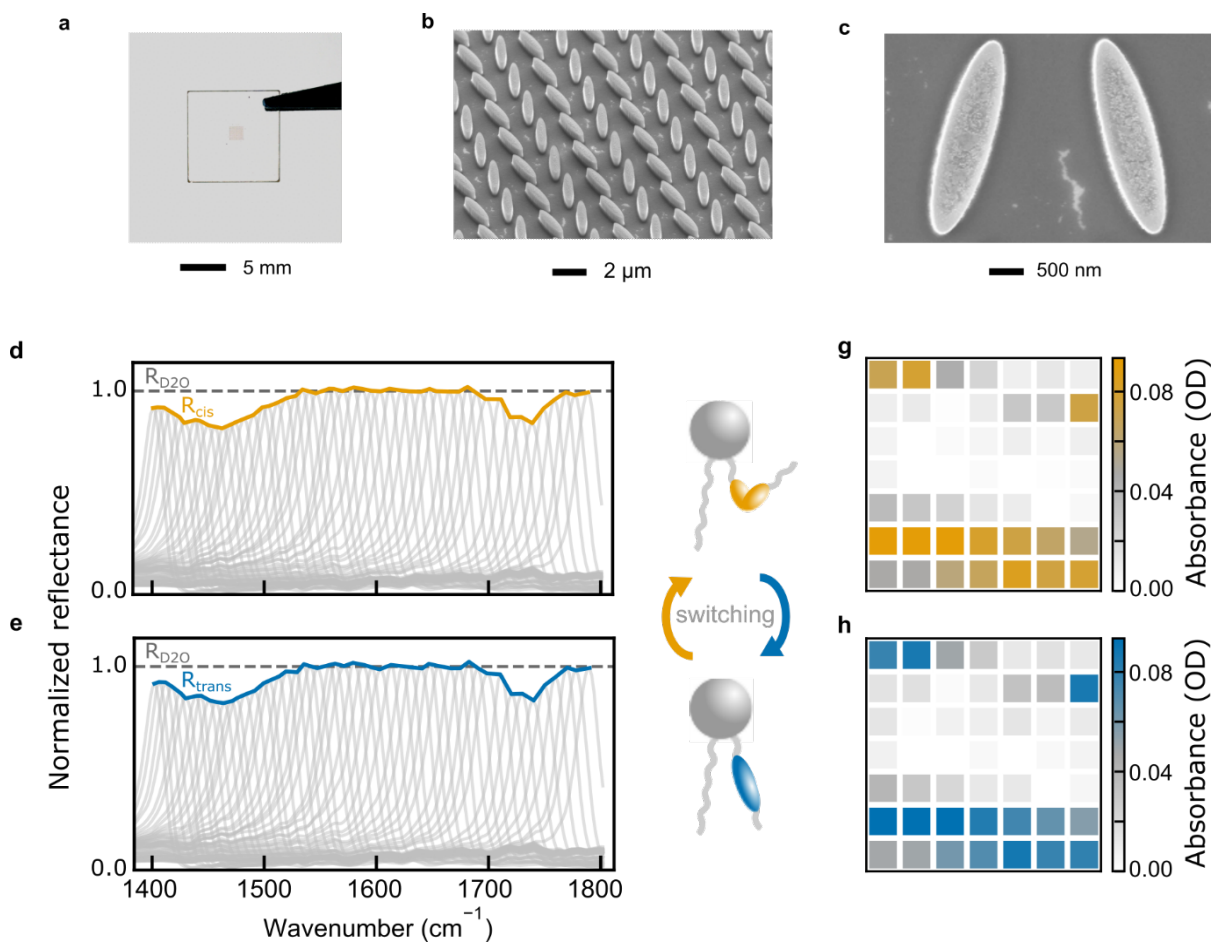

**Figure 3. Absorbance retrieval of lipid bilayers and spatial absorption mapping.** (**a**) Micrograph of metasurface on CaF2 substrate. (**b, c**) SEM images of metasurface unit cells. (**d,e**) Reflectance spectra for lipid bilayer in *cis* (d) and *trans* (e) conformation. (**f,g**) Absorbance spectra translated into 2D absorption map in reduced barcode scheme for *cis* (f) and *trans* (g) lipid bilayers.

To enable in-situ measurements, the fabricated metasurface chips were placed in a PDMS microfluidic chip connected to a syringe pump for flow-based analyte delivery. Spectroscopic measurements were performed in a laser-based infrared imaging microscope (for details see Methods). The setup utilizes stepwise scanning of the laser emission wavelength to measure the full spectrally resolved reflectance signal from all metapixels simultaneously. Continuous acquisition of this data during the switching experiments delivers a rich hyperspectral data set ideally suited for subsequent analysis using machine learning algorithms. To minimize attenuation due to the optical path through the aqueous medium, the reflectance was measured from the backside of the substrate, where a cutout in the microfluidic cell similarly avoids the IR absorption associated with PDMS.

For the membrane measurements, a previously reported tip-sonication protocol[30] was adapted to form small unilamellar vesicles (SUVs) of only AzoPC in $D_2O$ (See Materials and Methods). Subsequently 400 μL of the vesicle solution was injected into the microfluidic cell, allowing a surface-supported membrane to form via vesicle fusion. Metapixel reflectance spectra of a fully formed AzoPC membrane in the *cis* state are shown in Fig. 3d, clearly revealing the characteristic absorption signature of the lipids, with pronounced features at the characteristic wavenumbers associated with the $CH_2$ scissoring (1470 cm$^{-1}$) and C=O stretching (1735 cm$^{-1}$)[37]. Translated to an imaging-based representation in the reduced barcode scheme with 49 distinct pixels, these bands are visible as characteristic high-intensity regions at the top and bottom of the molecular barcode (Fig. 3f). Illumination of the membrane with light from a visible-spectrum LED at a wavelength of 465 nm converts the lipids back to their *trans* conformation, which is observed as a slight change in the metapixel reflectance spectra[37] (Fig. 3e,g).

Our pixelated metasurface platform provides high surface enhancements over a wide spectral range, but interpreting the signal from the metapixels and thus reliably detecting and discriminating biomolecules with this approach can be challenging. It often requires thorough manual data processing and is very sensitive to various signal drifts. To make the detection of biomolecules with pixelated metasurfaces more efficient, it is not only advantageous to automate data processing, but also to automatically select appropriate metapixel signals tailored to the specific application. To address this, we developed an automated, AI-based feature selection framework that leverages a pre-trained one-dimensional convolutional neural network (1D CNN). We initially trained this model on raw spectral molecular data generated by the sensor platform. The dataset, split into 70% training data and 30% validation data, comprises individual reflectance spectra of the 49 averaged metapixels, each labeled as *cis* or *trans* configuration (Fig. 4a).

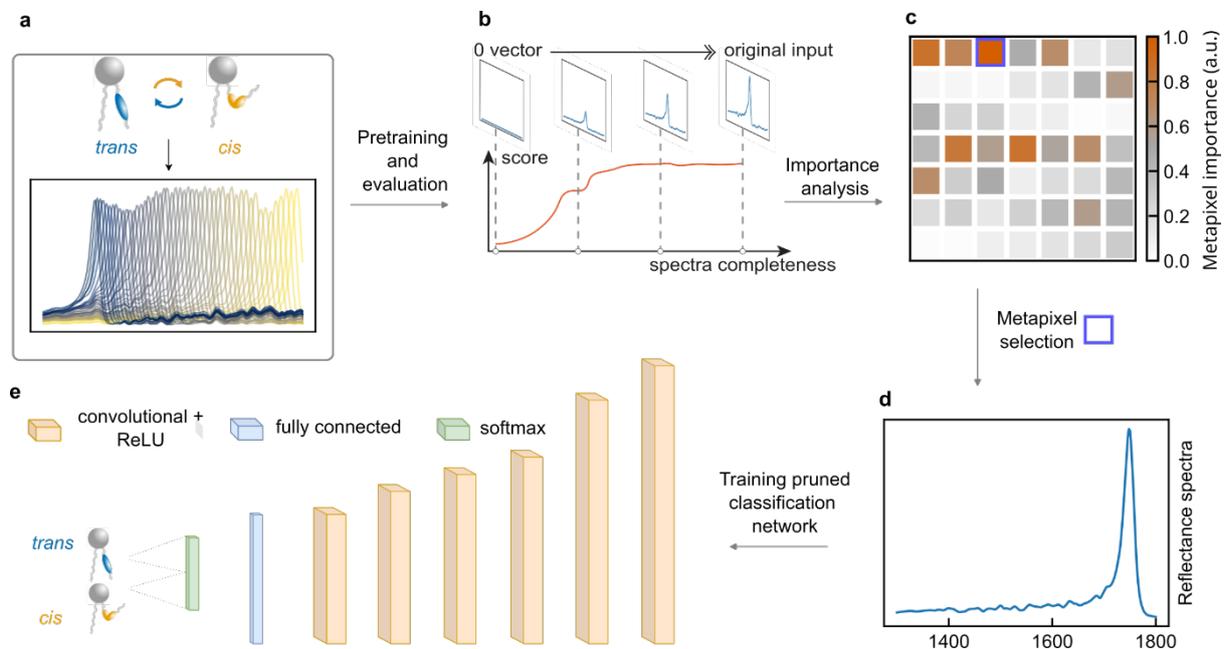

**Figure 4. Feature extraction framework and classification model.** (**a**) Sketch of hyperspectral data cube. (**b**) Sketch of the integrated gradient method. (**c**) Calculated importance score over each metapixel. (**d**) Reflection spectra of the selected metapixel. (**e**) Sketch of the pruned classification network.

After initial training, we used integrated gradients (IG), an explainable AI technique for the importance evaluation of all metapixels. This approach is based on backpropagation, calculating the gradient of each spectral input regarding the classification results, which represent the contribution of the network decisions[49]. As seen in Figure 4b, the IG computes an attribution score by accumulating the gradients calculated from a series of varying spectral inputs, starting from a baseline 0-vector and ending with the complete measured metapixel spectra. Such a spectral sweep ensures the sensitivity of the attribution process changes with the amplitude of the spectra[50]. The attribution plot (Fig. 4c) illustrates the overall importance of each metapixel concerning the given *cis* or *trans* classification task which is assigned to the pixels by the classification model during the decision process. The pixel with the highest attribution score (reflectance spectra in Fig. 4d) has a resonance that can be attributed to the C=O stretching of the anhydrous esters at 1742 cm$^{-1}$ [37]

Following feature selection, we transitioned to the refinement of our initial Convolutional Neural Network (CNN). We employed a process known as pruning (Fig. 4e), which aims to simplify the model by minimizing its complexity. This process is vital as it can lead to a model that is more computationally efficient and less prone to overfitting, without compromising performance. Remarkably, we achieved a 98% reduction in complexity, retaining only 251 out of the original parameters. The resulting pruned CNN, despite its streamlined structure, continued to process the reflection spectra of the selected metapixels effectively, outputting classifications into *trans* and *cis* categories. To validate the effectiveness of our pruned CNN, we turned to an analytical tool called the confusion matrix, which is depicted in Fig. 5a. In essence, a confusion matrix is a table that displays the performance of a supervised learning model. It's organized such that each row corresponds to the model's predictions for a given class, while each column shows the actual instances of that class. The intersection of a row and a column reveals the number of instances where the model predicted a particular class, and the true class was the same. This allows us to see not only where the model was correct, but also where and how it was incorrect. Despite our significant reduction in parameters, the confusion matrix reveals that our pruned CNN continues to perform exceptionally well.

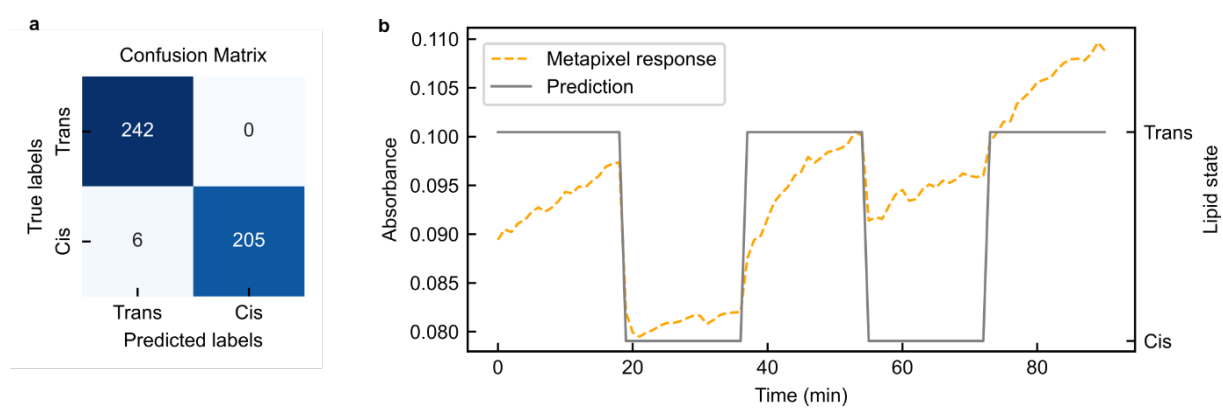

**Figure 5. Machine learning performance and prediction.** (**a**) Confusion matrix of the classification model. (**b**) Visualization of the model prediction vs. the absorbance of selected metapixel.

We then applied our methodology to the continuous monitoring of the photoswitching process of the lipids using time-series measurements. This method enabled us to detect and interpret multiple reversible conformational changes, which were triggered by UV/VIS light illumination. Our dynamic measurements commenced with a supported bilayer composed of AzoPC lipids in the *trans* state. Figure 5b illustrates the absorbance of the selected metapixel and the corresponding prediction by the pruned CNN. Remarkably, the pruned CNN's prediction aligned well with the observed switching cycles of approximately 18 minutes. This successful prediction of dynamic molecular events underscores the real-world utility and robustness of our pruned CNN. Despite its reduced complexity, the pruned CNN maintained a high level of accuracy, further establishing its effectiveness and potential for broader applications in the field of nanophotonics.

**Discussion**

We have introduced an integrated platform combining metasurfaces, optofluidics and deep learning which allows probing the structure and dynamics of biological entities in an aqueous environment in the mid-infrared. By specifically engineering metasurfaces supporting BICs, we realized the in-situ real-time investigation of the composition and switching dynamics of photolipid membranes with a rapid and reliable classification of the molecular measurements by the implemented convolutional neural network. Previous SEIRA implementations for studying molecular dynamics in-situ were mainly based on plasmonic resonators, which were shown to suffer from Ohmic losses reducing the sensitivity of the measurement[51,52]. More sensitive approaches based on all-dielectric metasurfaces supporting high-Q BIC resonances have so far mainly been applied in dry conditions or in the visible to near-infrared spectral range[22,24,53]. We overcome these constraints by engineering all-dielectric metasurfaces that give rise to high-Q BIC resonances also in aqueous environment in the mid-infrared and by this boost the capabilities of studying molecular changes with minute spectral variations in the molecular fingerprint region.

Accounting for the complexity of the data analysis of metasurface-enhanced bioanalytical studies, implementing deep learning models has shown great potential[23,52]. Therefore, we combined the neural network with a feature selection framework supported by integrated gradients (IG), a state-of-the-art explainable AI technique, that allowed us to automatically process our data and to reduce the number parameters by 98%. Overall, our system overcomes current limitations in surface-enhanced biospectroscopy, first by applying ultrasensitive high-Q BIC-supporting metasurfaces for in-situ studies in the mid-infrared, and second, by implementing deep learning for rapid and reliable data processing and classification. The presented machine learning model is a very promising tool not only for improved data analysis, but also for the optimization of metasurface designs. In this way, our integrated platform can be straightforwardly tailored to other molecular analytes by modifying the metasurface based on the CNN model results, paving the way for studies of more complex systems and processes ranging from cell signaling and chemotaxis in biology to air pollution and aerosol dispersion in environmental monitoring.

**Materials and Methods**

*Metasurface fabrication*

Silicon metasurfaces were nanofabricated on a $CaF_2$ substrate by electron beam lithography. Prior to fabrication, the $CaF_2$ substrates were cleaned and a 700nm thick amorphous silicon (a-Si) layer was deposited at 180 °C via plasma-enhanced chemical vapor deposition (PlasmaPro 100 PECVD, Oxford Instruments, United Kingdom). A layer of positive tone resist consisting of Poly(methyl mathacrylate) (PMMA) with a molecular weight of 950k was spin coated on the sample at 3000 rpm for 60 s and then pre-baked for three minutes at 180 °C. To avoid charging effects during electron beam exposure, a final layer of conductive polymer (Espacer 300Z) was spin-coated at 2000 rpm for 60 s. The designed structures were patterned into the sample by electron beam lithography (eLINE Plus, Raith GmbH, Germany). The patterning process was carried out at 20 kV acceleration voltage using an aperture size of 15 μm. The exposed resist was developed in a solution of isopropyl alcohol (IPA) and ultrapure water with a ratio of 7:3 for 60s at room temperature. As a hard mask, we used 20nm $SiO_2$ and 40nm Cr deposited via electron beam evaporation. The liftoff of the metal structures was performed in special remover (Microsposit remover 1165) at 80°C for 1-2h. The hard mask pattern was transferred into the silicon film via reactive ion etching (PlasmaPro 100 Cobra, Oxford Instruments, United Kingdom). Finally, the remaining hard mask was removed. For the chromium layer, a wet etchant (Cr etch 210, NB Technologies GmbH, Germany) was employed. The $SiO_2$ layer was removed with reactive ion etching, to obtain pure silicon nanostructures.

*Numerical simulations*

Numerical simulations were performed using the finite-element frequency-domain Maxwell solver included in CST Studio Suite 2020 (Dassault Systèmes). Reflectance and transmittance spectra were simulated under linearly polarized (TM) normally incident illumination using periodic Floquet boundary conditions. An open port facilitated the introduction of light. This port was positioned to introduce light through an air boundary (refractive index n = 1), while the same port on the opposite side transmitted the resulting power. Reflectance was quantified by comparing the reflected power to the introduced power.

*Photolipid vesicle preparation*

Small unilamellar photolipid vesicles were prepared by tip sonication as reported previously[30]: 100 µL of AzoPC lipids (c = 6.36 mM in $CHCl_3$ (amylene stabilized, Merck)) were dried using pressure air. After rehydration in 1.5 mL $D_2O$, the solution was tip-sonicated (Bandelin, Sonopuls) on ice twice for 30 s. The sample was then centrifuged with a relative centrifugal force of 35.8 $rpm^2$m and stored at 4°C until further use.

*Measurement setup*

Spectroscopic measurements were performed using a Spero microscope (Daylight Solutions Inc., USA) equipped with a 4x, 0.3 NA objective lens, providing a 2 x 2 $mm^2$ field of view. Infrared light from a quantum cascade laser module was linearly polarized and collected by an uncooled microbolometer focal plane array with 480 x 480 pixels. In reflectance mode, spectra were obtained in the range of 948 $cm^{-1}$ to 1800 $cm^{-1}$, with a spectral resolution of 2 $cm^{-1}$. Hyperspectral cubes were acquired continuously using the ChemVision software (Daylight Solutions Inc., USA), with each cube being captured and stored in 64 seconds. Background measurements were performed on a gold mirror before the start of each measurement. To drive the photoswitching, LEDs with the required center wavelengths were incorporated into the sensing platform. The metasurface chip was mounted upside down in a polydimethylsiloxane (PDMS) microfluidic cell, allowing the measurement of metapixels through the backside of the substrate. The microfluidic cell used in this study comprises an inlet and an outlet, interconnected by a channel measuring 150 µm in height and 500 µm in width. Positioned at the center of the channel, where the metasurface chip is affixed, a square reservoir measuring 7.5 x 7.5 mm² is located. A syringe pump was used to control the flow of the sample solution inside the cell, with a maximum flow rate of 500 µL $min^{-1}$. The pump was turned off during all spectroscopic measurements. The reflectance spectra were internally background-subtracted

ChemVision. In-house python code was then used to extract the metapixel spectra from the hyperspectral image data.

*Time-series experiments and absorbance calculation*

The microfluidic cell was filled with deuterium oxide ($D_2O$) and measured in the microscope continuously for 11 min. The $D_2O$ was exchanged with the vesicle solution containing the AzoPC lipids in either *trans* or *cis* state within 1 min. Immediately after, a continuous measurement was performed for 59 min for membrane formation by vesicle fusion. To switch the lipids in the opposite state, the corresponding LED was turned on for 4 min 30 s followed by turning off the LED for 15 min. These two steps were repeated three times. The absorbance in Fig. 5b was calculated via $A = -\log{(R_{\text{trans,cis}}/R_{\text{D2O}})}$ at the resonance position of the metapixel. Metapixel spectra for *cis* and *trans* states were recovered by averaging over the spectra from time series measurements with the AzoPC in the corresponding state. All data 14 minutes prior to the first LED illumination were excluded, as were data with the switching LEDs turned on. To recover the full absorbance spectrum, the calculated absorbances for the *cis* and *trans* states of metapixel spectra were linearly interpolated and smoothed by a third order Savitzky-Golay filter.

*1D CNN architecture and integrated gradients*

The architecture of our 1D convolutional neural network (CNN) model was implemented using PyTorch. The model consists of 6 convolutional layers, with a rectified linear unit (ReLU) being applied after every of the first 5 layers, followed by a fully connected final layer with Softmax activation. The use of ReLU functions increases the non-linearity of our model, while the Softmax function allocates probability values to the classification of the *trans* or *cis* state. The initial convolutional layer of the network for pretraining has 128 filters with a kernel size of 5x1 and stride of 2, while the second layer consists of 100 filters with a kernel size of 3x1 and stride of 1. The remaining layers have 80, 60, 40, and 20 filters, respectively, all with kernel sizes of 1x1 and stride of 1. The input layer consists of 50 averaged metapixel spectra from 1300 cm$^{-1}$ to 1800 cm$^{-1}$, with a step size of 2 cm$^{-1}$ (total 251 steps). Our training and validation dataset was built from 494 such spectra, obtained from measurements of AzoPC lipids in their two configurations, and labeled with the corresponding lipid state. The data was split 70/30 for training and validation. The network was trained for 160 epochs, and a dropout factor of 0.2 was used to reduce the risk of overfitting. The resulting model achieved an accuracy of 98%. The output layer is a 1x2 vector that classifies the input spectra into the *trans* and *cis* states. To quantitatively determine the contribution of each input feature to the output, we implemented the Integrated Gradients algorithm[49] in our architecture. We utilized the IntegratedGradients class from the Captum library[54] to calculate the attribution score for each input feature. Our inputs for the feature selection framework consisted of 50 averaged metapixel spectra, which were processed by a 1D convolutional neural network (CNN) that we had previously trained. The output labels were used to evaluate the performance of the model. To calculate the integrated gradients for a specific input feature, we first defined a baseline input, which was a vector of zeros. Then, for each point along the path from the baseline input to the actual input, we calculated the partial derivative of the model's output with respect to the input feature. Finally, we used Riemann integration to integrate these derivatives and obtain the attribution score for the input feature. After the initial training and the feature selection, we implemented a simplified version of the 1D CNN, which inputs one selected metapixel spectrum. The network contains four convolutional layers with filter sizes of 8,16,32,32 and kernel sizes of 7x1,5x1,5x1,5x1, respectively. We allocated a fully connected final layer with Softmax activation for the final prediction.

**Acknowledgements**

This project was funded by the Deutsche Forschungsgemeinschaft (DFG, German Research Foundation) under grant numbers EXC 2089/1–390776260 (Germany's Excellence Strategy), TI 1063/1 (Emmy Noether Program), and the Collaborative Research Center - SFB1032 (Project No. 201269156, project A8). We further acknowledge the Bavarian program Solar Energies Go Hybrid (SolTech) and the Center for NanoScience (CeNS). Funded by the European Union (ERC, METANEXT, 101078018 and NEHO, 101046329). Views and opinions expressed are however those of the author(s) only and do not necessarily reflect those of the European Union or the European Research Council Executive Agency. Neither the European Union nor the granting authority can be held responsible for them. S.A.M. additionally acknowledges the Lee-Lucas Chair in Physics and the EPSRC (EP/W017075/1) and S.D.P acknowledges the European Research Council Consolidator Grant "ProForce".

**Data availability**

All data needed to evaluate the conclusions in the paper are present in the paper and/or the Supplementary Materials.

**References**

(1) Stuart, B. H. *Infrared Spectroscopy: Fundamentals and Applications*; John Wiley & Sons, 2004.

(2) Braiman, M. S.; Rothschild, K. J. Fourier Transform Infrared Techniques for Probing Membrane Protein Structure. *Annual Review of Biophysics and Biophysical Chemistry* **1988**, *17* (1), 541–570. https://doi.org/10.1146/annurev.bb.17.060188.002545.

(3) Khalil, M.; Demirdöven, N.; Tokmakoff, A. Coherent 2D IR Spectroscopy:  Molecular Structure and Dynamics in Solution. *J. Phys. Chem. A* **2003**, *107* (27), 5258–5279. https://doi.org/10.1021/jp0219247.

(4) Neubrech, F.; Huck, C.; Weber, K.; Pucci, A.; Giessen, H. Surface-Enhanced Infrared Spectroscopy Using Resonant Nanoantennas. *Chem. Rev.* **2017**, *117* (7), 5110–5145. https://doi.org/10.1021/acs.chemrev.6b00743.

(5) Altug, H.; Oh, S.-H.; Maier, S. A.; Homola, J. Advances and Applications of Nanophotonic Biosensors. *Nat. Nanotechnol.* **2022**, *17* (1), 5–16. https://doi.org/10.1038/s41565-021-01045-5.

(6) Dong, L.; Yang, X.; Zhang, C.; Cerjan, B.; Zhou, L.; Tseng, M. L.; Zhang, Y.; Alabastri, A.; Nordlander, P.; Halas, N. J. Nanogapped Au Antennas for Ultrasensitive Surface-Enhanced Infrared Absorption Spectroscopy. *Nano Lett.* **2017**, *17* (9), 5768–5774. https://doi.org/10.1021/acs.nanolett.7b02736.

(7) Chen, P.; Chung, M. T.; McHugh, W.; Nidetz, R.; Li, Y.; Fu, J.; Cornell, T. T.; Shanley, T. P.; Kurabayashi, K. Multiplex Serum Cytokine Immunoassay Using Nanoplasmonic Biosensor Microarrays. *ACS Nano* **2015**, *9* (4), 4173–4181. https://doi.org/10.1021/acsnano.5b00396.

(8) Yang, X.; Sun, Z.; Low, T.; Hu, H.; Guo, X.; Abajo, F. J. G. de; Avouris, P.; Dai, Q. Nanomaterial-Based Plasmon-Enhanced Infrared Spectroscopy. *Advanced Materials* **2018**, *30* (20), 1704896. https://doi.org/10.1002/adma.201704896.

(9) Yavas, O.; Aćimović, S. S.; Garcia-Guirado, J.; Berthelot, J.; Dobosz, P.; Sanz, V.; Quidant, R. Self-Calibrating On-Chip Localized Surface Plasmon Resonance Sensing for Quantitative and Multiplexed Detection of Cancer Markers in Human Serum. *ACS Sens.* **2018**, *3* (7), 1376–1384. https://doi.org/10.1021/acssensors.8b00305.

(10) John-Herpin, A.; Tittl, A.; Kühner, L.; Richter, F.; Huang, S. H.; Shvets, G.; Oh, S.-H.; Altug, H. Metasurface-Enhanced Infrared Spectroscopy: An Abundance of Materials and Functionalities. *Advanced Materials n/a* (n/a), 2110163. https://doi.org/10.1002/adma.202110163.

(11) Tseng, M. L.; Jahani, Y.; Leitis, A.; Altug, H. Dielectric Metasurfaces Enabling Advanced Optical Biosensors. *ACS Photonics* **2021**, *8* (1), 47–60. https://doi.org/10.1021/acsphotonics.0c01030.

(12) Krasnok, A.; Caldarola, M.; Bonod, N.; Alú, A. Spectroscopy and Biosensing with Optically Resonant Dielectric Nanostructures. *Advanced Optical Materials* **2018**, *6* (5), 1701094. https://doi.org/10.1002/adom.201701094.

(13) Yavas, O.; Svedendahl, M.; Dobosz, P.; Sanz, V.; Quidant, R. On-a-Chip Biosensing Based on All-Dielectric Nanoresonators. *Nano Lett.* **2017**, *17* (7), 4421–4426. https://doi.org/10.1021/acs.nanolett.7b01518.

(14) Koshelev, K.; Favraud, G.; Bogdanov, A.; Kivshar, Y.; Fratalocchi, A. Nonradiating Photonics with Resonant Dielectric Nanostructures. *Nanophotonics* **2019**, *8* (5), 725–745. https://doi.org/10.1515/nanoph-2019-0024.

(15) Koshelev, K.; Lepeshov, S.; Liu, M.; Bogdanov, A.; Kivshar, Y. Asymmetric Metasurfaces with High-$Q$ Resonances Governed by Bound States in the Continuum. *Phys. Rev. Lett.* **2018**, *121* (19), 193903. https://doi.org/10.1103/PhysRevLett.121.193903.

(16) Hsu, C. W.; Zhen, B.; Stone, A. D.; Joannopoulos, J. D.; Soljačić, M. Bound States in the Continuum. *Nature Reviews Materials* **2016**, *1* (9), 1–13. https://doi.org/10.1038/natrevmats.2016.48.

(17) Rybin, M. V.; Koshelev, K. L.; Sadrieva, Z. F.; Samusev, K. B.; Bogdanov, A. A.; Limonov, M. F.; Kivshar, Y. S. High-$Q$ Supercavity Modes in Subwavelength Dielectric Resonators. *Phys. Rev. Lett.* **2017**, *119* (24), 243901. https://doi.org/10.1103/PhysRevLett.119.243901.

(18) Ha, S. T.; Fu, Y. H.; Emani, N. K.; Pan, Z.; Bakker, R. M.; Paniagua-Domínguez, R.; Kuznetsov, A. I. Directional Lasing in Resonant Semiconductor Nanoantenna Arrays. *Nature Nanotech* **2018**, *13* (11), 1042–1047. https://doi.org/10.1038/s41565-018-0245-5.

(19) Li, S.; Zhou, C.; Liu, T.; Xiao, S. Symmetry-Protected Bound States in the Continuum Supported by All-Dielectric Metasurfaces. *Phys. Rev. A* **2019**, *100* (6), 063803. https://doi.org/10.1103/PhysRevA.100.063803.

(20) Leitis, A.; Tittl, A.; Liu, M.; Lee, B. H.; Gu, M. B.; Kivshar, Y. S.; Altug, H. Angle-Multiplexed All-Dielectric Metasurfaces for Broadband Molecular Fingerprint Retrieval. *Science Advances* **2019**, *5* (5), eaaw2871. https://doi.org/10.1126/sciadv.aaw2871.

(21) Yesilkoy, F.; Arvelo, E. R.; Jahani, Y.; Liu, M.; Tittl, A.; Cevher, V.; Kivshar, Y.; Altug, H. Ultrasensitive Hyperspectral Imaging and Biodetection Enabled by Dielectric Metasurfaces. *Nat. Photonics* **2019**, *13* (6), 390–396. https://doi.org/10.1038/s41566-019-0394-6.

(22) Jahani, Y.; Arvelo, E. R.; Yesilkoy, F.; Koshelev, K.; Cianciaruso, C.; De Palma, M.; Kivshar, Y.; Altug, H. Imaging-Based Spectrometer-Less Optofluidic Biosensors Based on Dielectric Metasurfaces for Detecting Extracellular Vesicles. *Nat Commun* **2021**, *12* (1), 3246. https://doi.org/10.1038/s41467-021-23257-y.

(23) Tittl, A.; John-Herpin, A.; Leitis, A.; Arvelo, E. R.; Altug, H. Metasurface-Based Molecular Biosensing Aided by Artificial Intelligence. *Angewandte Chemie*

*International Edition* **2019**, *58* (42), 14810–14822. https://doi.org/10.1002/anie.201901443.

(24) Tittl, A.; Leitis, A.; Liu, M.; Yesilkoy, F.; Choi, D.-Y.; Neshev, D. N.; Kivshar, Y. S.; Altug, H. Imaging-Based Molecular Barcoding with Pixelated Dielectric Metasurfaces. *Science* **2018**, *360* (6393), 1105–1109. https://doi.org/10.1126/science.aas9768.

(25) Coskun, Ü.; Simons, K. Cell Membranes: The Lipid Perspective. *Structure* **2011**, *19* (11), 1543–1548. https://doi.org/10.1016/j.str.2011.10.010.

(26) Chan, Y.-H. M.; Boxer, S. G. Model Membrane Systems and Their Applications. *Current Opinion in Chemical Biology* **2007**, *11* (6), 581–587. https://doi.org/10.1016/j.cbpa.2007.09.020.

(27) Khan, M. S.; Dosoky, N. S.; Williams, J. D. Engineering Lipid Bilayer Membranes for Protein Studies. *International Journal of Molecular Sciences* **2013**, *14* (11), 21561–21597. https://doi.org/10.3390/ijms141121561.

(28) Seu, K. J.; Cambrea, L. R.; Everly, R. M.; Hovis, J. S. Influence of Lipid Chemistry on Membrane Fluidity: Tail and Headgroup Interactions. *Biophysical Journal* **2006**, *91* (10), 3727–3735. https://doi.org/10.1529/biophysj.106.084590.

(29) Morstein, J.; Impastato, A. C.; Trauner, D. Photoswitchable Lipids. *Chembiochem* **2021**, *22* (1), 73–83. https://doi.org/10.1002/cbic.202000449.

(30) Urban, P.; Pritzl, S. D.; Ober, M. F.; Dirscherl, C. F.; Pernpeintner, C.; Konrad, D. B.; Frank, J. A.; Trauner, D.; Nickel, B.; Lohmueller, T. A Lipid Photoswitch Controls Fluidity in Supported Bilayer Membranes. *Langmuir* **2020**, *36* (10), 2629–2634. https://doi.org/10.1021/acs.langmuir.9b02942.

(31) Urban, P.; Pritzl, S. D.; Konrad, D. B.; Frank, J. A.; Pernpeintner, C.; Roeske, C. R.; Trauner, D.; Lohmüller, T. Light-Controlled Lipid Interaction and Membrane Organization in Photolipid Bilayer Vesicles. *Langmuir* **2018**, *34* (44), 13368–13374. https://doi.org/10.1021/acs.langmuir.8b03241.

(32) Pritzl, S. D.; Urban, P.; Prasselsperger, A.; Konrad, D. B.; Frank, J. A.; Trauner, D.; Lohmüller, T. Photolipid Bilayer Permeability Is Controlled by Transient Pore Formation. *Langmuir* **2020**, *36* (45), 13509–13515. https://doi.org/10.1021/acs.langmuir.0c02229.

(33) Pernpeintner, C.; Frank, J. A.; Urban, P.; Roeske, C. R.; Pritzl, S. D.; Trauner, D.; Lohmüller, T. Light-Controlled Membrane Mechanics and Shape Transitions of Photoswitchable Lipid Vesicles. *Langmuir* **2017**, *33* (16), 4083–4089. https://doi.org/10.1021/acs.langmuir.7b01020.

(34) Pritzl, S. D.; Morstein, J.; Kahler, S.; Konrad, D. B.; Trauner, D.; Lohmüller, T. Postsynthetic Photocontrol of Giant Liposomes via Fusion-Based Photolipid Doping. *Langmuir* **2022**, *38* (39), 11941–11949. https://doi.org/10.1021/acs.langmuir.2c01685.

(35) Kuiper, J. M.; Engberts, J. B. F. N. H-Aggregation of Azobenzene-Substituted Amphiphiles in Vesicular Membranes. *Langmuir* **2004**, *20* (4), 1152–1160. https://doi.org/10.1021/la0358724.

(36) Doroudgar, M.; Morstein, J.; Becker-Baldus, J.; Trauner, D.; Glaubitz, C. How Photoswitchable Lipids Affect the Order and Dynamics of Lipid Bilayers and Embedded Proteins. *J. Am. Chem. Soc.* **2021**, *143* (25), 9515–9528. https://doi.org/10.1021/jacs.1c03524.

(37) Crea, F.; Vorkas, A.; Redlich, A.; Cruz, R.; Shi, C.; Trauner, D.; Lange, A.; Schlesinger, R.; Heberle, J. Photoactivation of a Mechanosensitive Channel. *Frontiers in Molecular Biosciences* **2022**, *9*.

(38) Chander, N.; Morstein, J.; Bolten, J. S.; Shemet, A.; Cullis, P. R.; Trauner, D.; Witzigmann, D. Optimized Photoactivatable Lipid Nanoparticles Enable Red Light Triggered Drug Release. *Small* **2021**, *17* (21), 2008198. https://doi.org/10.1002/smll.202008198.

(39) Jiménez-Rojo, N.; Feng, S.; Morstein, J.; Pritzl, S. D.; Harayama, T.; Asaro, A.; Vepřek, N. A.; Arp, C. J.; Reynders, M.; Novak, A. J. E.; Kanshin, E.; Ueberheide, B.; Lohmüller, T.; Riezman, H.; Trauner, D. Optical Control of Membrane Fluidity Modulates Protein Secretion. *bioRxiv* **2022**, 2022.02.14.480333. https://doi.org/10.1101/2022.02.14.480333.

(40) Dluhy, R. A. Infrared Spectroscopy of Biophysical Monomolecular Films at Interfaces: Theory and Applications. *Applied Spectroscopy Reviews* **2000**, *35* (4), 315–351. https://doi.org/10.1081/ASR-100101228.

(41) Backus, E. H. G.; Kuiper, J. M.; Engberts, J. B. F. N.; Poolman, B.; Bonn, M. Reversible Optical Control of Monolayers on Water through Photoswitchable Lipids. *J. Phys. Chem. B* **2011**, *115* (10), 2294–2302. https://doi.org/10.1021/jp1113619.

(42) Cernescu, A.; Szuwarzyński, M.; Kwolek, U.; Wydro, P.; Kepczynski, M.; Zapotoczny, S.; Nowakowska, M.; Quaroni, L. Label-Free Infrared Spectroscopy and Imaging of Single Phospholipid Bilayers with Nanoscale Resolution. *Anal. Chem.* **2018**, *90* (17), 10179–10186. https://doi.org/10.1021/acs.analchem.8b00485.

(43) Adão, T.; Hruška, J.; Pádua, L.; Bessa, J.; Peres, E.; Morais, R.; Sousa, J. J. Hyperspectral Imaging: A Review on UAV-Based Sensors, Data Processing and Applications for Agriculture and Forestry. *Remote Sensing* **2017**, *9* (11), 1110. https://doi.org/10.3390/rs9111110.

(44) He, H.; Yan, S.; Lyu, D.; Xu, M.; Ye, R.; Zheng, P.; Lu, X.; Wang, L.; Ren, B. Deep Learning for Biospectroscopy and Biospectral Imaging: State-of-the-Art and Perspectives. *Anal. Chem.* **2021**, *93* (8), 3653–3665. https://doi.org/10.1021/acs.analchem.0c04671.

(45) Huang, L.; Luo, R.; Liu, X.; Hao, X. Spectral Imaging with Deep Learning. *Light Sci Appl* **2022**, *11* (1), 61. https://doi.org/10.1038/s41377-022-00743-6.

(46) Molnar, C. *Interpretable Machine Learning*. Christoph Molnar. https://christophmolnar.com/books/interpretable-machine-learning/ (accessed 2023-06-20).

(47) De Lucia, G.; Lapegna, M.; Romano, D. Towards Explainable AI for Hyperspectral Image Classification in Edge Computing Environments. *Comput. Electr. Eng.* **2022**, *103* (C). https://doi.org/10.1016/j.compeleceng.2022.108381.

(48) Kiranyaz, S.; Avci, O.; Abdeljaber, O.; Ince, T.; Gabbouj, M.; Inman, D. J. 1D Convolutional Neural Networks and Applications: A Survey. *Mechanical Systems and Signal Processing* **2021**, *151*, 107398. https://doi.org/10.1016/j.ymssp.2020.107398.

(49) Sundararajan, M.; Taly, A.; Yan, Q. Axiomatic Attribution for Deep Networks. **2017**. https://doi.org/10.48550/ARXIV.1703.01365.

(50) Miglani, V.; Kokhlikyan, N.; Alsallakh, B.; Martin, M.; Reblitz-Richardson, O. Investigating Saturation Effects in Integrated Gradients. *arXiv preprint* **2020**. https://doi.org/10.48550/arXiv.2010.12697.

(51) Rodrigo, D.; Tittl, A.; Ait-Bouziad, N.; John-Herpin, A.; Limaj, O.; Kelly, C.; Yoo, D.; Wittenberg, N. J.; Oh, S.-H.; Lashuel, H. A.; Altug, H. Resolving Molecule-Specific Information in Dynamic Lipid Membrane Processes with Multi-Resonant Infrared Metasurfaces. *Nature Communications* **2018**, *9* (1), 2160. https://doi.org/10.1038/s41467-018-04594-x.

(52) John-Herpin, A.; Kavungal, D.; von Mücke, L.; Altug, H. Infrared Metasurface Augmented by Deep Learning for Monitoring Dynamics between All Major Classes of Biomolecules. *Advanced Materials* **2021**, *33* (14), 2006054. https://doi.org/10.1002/adma.202006054.

(53) Wang, Y.; Ali, Md. A.; Chow, E. K. C.; Dong, L.; Lu, M. An Optofluidic Metasurface for Lateral Flow-through Detection of Breast Cancer Biomarker. *Biosensors and Bioelectronics* **2018**, *107*, 224–229. https://doi.org/10.1016/j.bios.2018.02.038.

(54) Kokhlikyan, N.; Miglani, V.; Martin, M.; Wang, E.; Alsallakh, B.; Reynolds, J.; Melnikov, A.; Kliushkina, N.; Araya, C.; Yan, S.; Reblitz-Richardson, O. Captum: A Unified and Generic Model Interpretability Library for PyTorch. **2020**. https://doi.org/10.48550/ARXIV.2009.07896.